\documentclass[aps,prb,twocolumn,groupedaddress,amsmath]{revtex4-2}
\usepackage{graphicx}
\usepackage{amssymb}
\usepackage[normalem]{ulem}
\usepackage{MnSymbol}
\usepackage{stmaryrd}
\usepackage{threeparttable}
\begin{document}
%\newcommand{\vn}[1]{{\bf{#1}}}
%\maxdeadcycles=1000
%\setcounter{topnumber}{8}
%\setcounter{bottomnumber}{8}
%\setcounter{totalnumber}{8}
\newcommand{\magdir}{\hat{\vn{n}}}
\newcommand{\vn}[1]{{\boldsymbol{#1}}}
\newcommand{\vht}[1]{{\boldsymbol{#1}}}
\newcommand{\polarivec}{\boldsymbol{\epsilon}}
\newcommand{\matn}[1]{{\bf{#1}}}
\newcommand{\matnht}[1]{{\boldsymbol{#1}}}
\newcommand{\bege}{\begin{equation}}
\newcommand{\ee}{\end{equation}}
\newcommand{\bal}{\begin{aligned}}
\newcommand{\defbar}{\overline}
\newcommand{\SM}{\scriptstyle}
\newcommand{\eal}{\end{aligned}}
\newcommand{\torkance}{t}
\newcommand{\udot}{\overset{.}{u}}
\newcommand{\exponential}[1]{{\exp(#1)}}
\newcommand{\phandot}[1]{\overset{\phantom{.}}{#1}}
\newcommand{\phandag}{\phantom{\dagger}}
\newcommand{\Trace}{\text{Tr}}
\newcommand{\Bxc}{\Omega}
\newcommand{\mubo}{\mu_{\rm B}^{\phantom{B}}}
\newcommand{\rmd}{{\rm d}}
\newcommand{\rme}{{\rm e}}
\newcommand{\intkspa}{\int\!\!\frac{\rmd^3 k}{(2\pi)^3}}
\newcommand{\intkspatwodim}{\int\!\!\frac{\rmd^2 k}{(2\pi)^2}}
\setcounter{secnumdepth}{2}
\title{Laser-induced torques in metallic antiferromagnets}
\author{Frank Freimuth$^{1,2}$}
\email[Corresp.~author:~]{f.freimuth@fz-juelich.de}
\author{Stefan Bl\"ugel$^{1}$}
\author{Yuriy Mokrousov$^{1,2}$}
\affiliation{$^1$Peter Gr\"unberg Institut and Institute for Advanced Simulation,
Forschungszentrum J\"ulich and JARA, 52425 J\"ulich, Germany}
\affiliation{$^2$Institute of Physics, Johannes Gutenberg University Mainz, 55099 Mainz, Germany
}
\date{\today}
\begin{abstract}
We study the laser-induced torques in the antiferromagnet (AFM)
Mn$_2$Au. We find that even linearly polarized light
may induce laser-induced torques in Mn$_2$Au, i.e., the light
does not have to be circularly polarized.
The laser-induced torques in Mn$_2$Au are comparable in magnitude
to those in the ferromagnets Fe, Co and FePt at optical frequencies.
We also compute the laser-induced torques at terahertz (THz) frequencies
and compare them to the spin-orbit torques (SOTs) excited by THz
laser-pulses.
We find the SOTs to be dominant at THz frequencies for the laser-field
strengths used in experiments.
Additionally, we show that the matrix elements of the spin-orbit
interaction (SOI) can
be used to 
add SOI only during the Wannier interpolation, which we call 
Wannier interpolation of SOI (WISOI). This technique allows us to
perform the Wannier interpolation conveniently for many
magnetization directions from a single set of Wannier functions.

\end{abstract}

\maketitle
\section{Introduction}
Using femtosecond laser-pulses to
switch the
magnetization~\cite{Lambert_all_optical_control,magnetization_switching_FePt},
to exert torques on the 
magnetic 
moments~\cite{Kimel_ultrafast_control_magnetization,nemec_ostt,femtosecond_control_electric_currents_Huisman,Kampen_optical_probe_coherent_spin_waves,ultrafast_stt_laser,choi_thermal_stt,Rabi_CoFeB,optical_driven_magnetization_dynamics_choi_2017,lasintor},
to move domain walls~\cite{PhysRevB.101.174418}, and
to excite magnons~\cite{PhysRevB.101.174427}
are promising concepts to write, store and process information on
ultrafast timescales in prospective device applications.
In bulk crystals laser-induced torques on the magnetization are
attributed to the inverse Faraday effect (IFE)
and to the optical spin-transfer 
torque (OSTT)~\cite{Kimel_ultrafast_control_magnetization,nemec_ostt,lasintor}.
Phenomenology for non-magnets predicts the IFE only for circularly polarized light.
However, works on the ferromagnetic Rashba model~\cite{light_induced_magnetic_field_rashba}
as well as
first-principles calculations~\cite{lasintor,PhysRevLett.117.137203} 
show that the IFE in ferromagnets differs from these predictions,
i.e., the IFE is present even for linearly polarized light.

Due to their terahertz (THz) magnetization dynamics, antiferromagnets (AFMs)
are another promising ingredient in ultrafast magnetism 
concepts~\cite{afmreview_RMP_90_015005,rmp_sot,doi:10.1063/1.4862467,THz_writing_speed_afm}.
To this end, spin-orbit torques (SOTs) in the bulk AFM Mn$_2$Au have
been studied intensively both in
theory~\cite{PhysRevLett.113.157201,PhysRevB.95.014403,orbitally_dominated_ree}
and in
experiment~\cite{writing_reading_AFM_Mn2Au,PhysRevApplied.9.064040,PhysRevApplied.9.054028,Mn2Au_PhysRevB.99.140409}
and it has been shown that the SOT may be used to switch
the N\'eel vector. For the optical manipulation of the N\'eel vector the
IFE
and OSTT might be promising alternatives to the SOT.
However, theoretical works on the IFE and OSTT in  antiferromagnets (AFM) are
still lacking.

When lasers at optical frequencies are used to exert torques in AFMs
it is clear that the SOT excited by the electric field of the laser
may be ignored, because it is oscillating at the laser frequency,
which
is far above the magnetic resonances in the AFMs in the THz range.
This picture changes when THz lasers are used to excite the
magnetization in AFMs.
In Ref.~\cite{THz_writing_speed_afm}
THz laser pulses were used to switch the AFM CuMnAs
contactlessly and this switching was interpreted as the action of the
SOT. 
While the SOT is linear in the applied
 electric field, the IFE and OSTT are quadratic in it. The question
therefore arises at which electric field strength the IFE and OSTT
become more important than the SOT when the frequency of the
applied electric field is in the THz range.

In this work we investigate the laser-induced torques
in the bulk AFM Mn$_2$Au. Our computational approach
is based on the Keldysh nonequilibrium formalism
and on the Wannier interpolation~\cite{wannier90communitycode} of the electronic
structure obtained from realistic density-functional
theory calculations. In order to study the dependence
of the laser-induced torques on the magnetization direction
we introduce a method that allows us to do the Wannier
interpolation conveniently for many different magnetization
directions on the basis of a single set of maximally localized Wannier
functions (MLWFs),
which we call Wannier interpolation of SOI (WISOI).

This paper is organized as follows.
In Sec.~\ref{sec_keldysh_formalism}
we briefly review the formalism that we use to compute the
laser-induced torque.
In Sec.~\ref{sec_soi_matrix} we describe WISOI briefly, deferring
details
on the implementation to Appendix~\ref{sec_app_soi_matrix}. 
In Sec.~\ref{sec_symmetry} we discuss how the symmetry of the Mn$_2$Au
crystal determines the form of the response tensor.
In Sec.~\ref{sec_results} we discuss our results on the laser-induced
torques
in Mn$_2$Au.
This paper ends with a summary in Sec.~\ref{sec_summary}. 

\section{Formalism}
\subsection{Keldysh formalism}
\label{sec_keldysh_formalism}

In Ref.~\cite{lasintor} we derived the following expression for the
laser-induced torque based on the Keldysh nonequilibrium formalism:
\bege\label{eq_define_torque_chi}
T_{i}=\frac{a_{0}^3 I}{c}
\left(
\frac{\mathcal{E}_{\rm H}}{\hbar\omega}
\right)^2
{\rm Im}
\sum_{jk}
\epsilon_j
\epsilon_k^*
\chi_{ijk},
\ee
where $c$ is the velocity of light,
$a_{0}=4\pi\epsilon_0\hbar^2/(me^2)$ is Bohr's radius,
$I=\epsilon_0 c E_{0}^2/2$ is the intensity of light,
$\epsilon_0$ is the vacuum permittivity,
$\mathcal{E}_{\rm H}=e^2/(4\pi\epsilon_0 a_0)$ is the Hartree energy,
and $\epsilon_{j}$ is the $j$-th Cartesian component of the light 
polarization vector.
The tensor $\chi_{ijk}$ is given by
\bege\label{eq_chi_noabrev}
\begin{aligned}
\chi_{ijk}=&\frac{2}{\mathcal{N}\hbar a_0^2 \mathcal{E}_{\rm H}}\sum_{\vn{k}}
\int \rmd \mathcal{E}
{\rm Tr}\Big[\\
&f(\mathcal{E})
\mathcal{T}_{i}
G^{\rm R}_{\vn{k}}(\mathcal{E})
v_{j}
G^{\rm R}_{\vn{k}}(\mathcal{E}-\hbar\omega)
v_{k}
G^{\rm R}_{\vn{k}}(\mathcal{E})
\\
-&
f(\mathcal{E})
\mathcal{T}_{i}
G^{\rm R}_{\vn{k}}(\mathcal{E})
v_{j}
G^{\rm R}_{\vn{k}}(\mathcal{E}-\hbar\omega)
v_{k}
G^{\rm A}_{\vn{k}}(\mathcal{E})\\
+&f(\mathcal{E})
\mathcal{T}_{i}
G^{\rm R}_{\vn{k}}(\mathcal{E})
v_{k}
G^{\rm R}_{\vn{k}}(\mathcal{E}+\hbar\omega)
v_{j}
G^{\rm R}_{\vn{k}}(\mathcal{E})
\\
-&f(\mathcal{E})
\mathcal{T}_{i}
G^{\rm R}_{\vn{k}}(\mathcal{E})
v_{k}
G^{\rm R}_{\vn{k}}(\mathcal{E}+\hbar\omega)
v_{j}
G^{\rm A}_{\vn{k}}(\mathcal{E})
\\
+&
f(\mathcal{E}-\hbar\omega)
\mathcal{T}_{i}
G^{\rm R}_{\vn{k}}(\mathcal{E})
v_{j}
G^{\rm R}_{\vn{k}}(\mathcal{E}-\hbar\omega)
v_{k}
G^{\rm A}_{\vn{k}}(\mathcal{E})\\
+&
f(\mathcal{E}+\hbar\omega)
\mathcal{T}_{i}
G^{\rm R}_{\vn{k}}(\mathcal{E})
v_{k}
G^{\rm R}_{\vn{k}}(\mathcal{E}+\hbar\omega)
v_{j}
G^{\rm A}_{\vn{k}}(\mathcal{E})
\Big],\\
\end{aligned}
\ee
where $\mathcal{N}$ is the number of $\vn{k}$ points, $f(\mathcal{E})$ 
is the Fermi distribution function, $\vn{\mathcal{T}}$ is the torque
operator, $\vn{v}$ is the velocity operator,
\bege \label{eq_define_green_analy}
G^{\rm R}_{\vn{k}}(\mathcal{E})=
\hbar
\sum_{n}\frac{|\vn{k}n\rangle\langle\vn{k}n|}{\mathcal{E}-\mathcal{E}_{\vn{k}n}+i\Gamma},
\ee
is the retarded Green function and
$G^{\rm A}_{\vn{k}}(\mathcal{E})=[G^{\rm R}_{\vn{k}}(\mathcal{E})]^{\dagger}$ is the advanced Green function.
Here, $|\vn{k}n\rangle$ and $\mathcal{E}_{\vn{k}n}$ 
are eigenstates and eigenenergies, respectively, 
and $\Gamma$ is a  constant broadening used
to simulate disorder and finite lifetimes of the electronic
states.

Our Green's function expressions provide an alternative approach
to the expressions based on the order
formalism~\cite{PhysRevB.85.045117,PhysRevLett.117.137203}
used to compute the IFE.
In Ref.~\cite{lasincucspira} we have shown that
Eq.~\eqref{eq_chi_noabrev}
may be modified in order to compute laser-induced charge and spin
photocurrents. Thus, the Keldysh formalism underlying Eq.~\eqref{eq_chi_noabrev}
provides also an alternative approach to the method based on the expressions
given first by von Baltz and Kraut~\cite{PhysRevB.19.1548,PhysRevB.23.5590,PhysRevB.97.241118}.

In the case of the SOT one may distinguish spin and 
orbital
torques~\cite{PhysRevResearch.2.033401,PhysRevLett.125.177201,tazaki2020currentinduced,PhysRevB.103.L020407,orbitally_dominated_ree,PhysRevResearch.3.013275}. For
example in TmIG/Pt/CuO$_{x}$ 
the SOT has been interpreted in terms of an orbital current
generated at the Pt/CuO$_{x}$ interface, which is converted
subsequently into a spin current by the SOI of
Pt~\cite{PhysRevLett.125.177201},
and in Ni/W(110) calculations show that the SOT even differs
in sign from the spin-transfer torque associated with the spin
current of W~\cite{PhysRevResearch.2.033401} because the orbital
torque dominates. However, when one computes the linear response
of the torque operator $\vn{\mathcal{T}}$ to the applied 
electric field one captures both
the spin torque and the orbital
torque.
Therefore, in order to describe the SOT when the orbital torque
is dominating, one does not need a new formalism to compute the
total SOT, but one might wish to develop additional tools to separate the total SOT into
spin and orbital contributions~\cite{PhysRevResearch.2.033401}.  

In the case of the laser-induced torques it might become useful
as well to distinguish between spin and orbital contributions.
A strong indication that this might be the case is the finding of a
large laser-induced orbital magnetization~\cite{PhysRevLett.117.137203} even without SOI.
However, the only mechanism by which this orbital magnetization may
directly lead to a torque on the magnetization is through the Oersted
field generated by the orbital photocurrent. Therefore, one may expect
that a significantly larger torque arises from the coupling of the
laser-induced orbital magnetization to the spin magnetization due to
SOI. Such an indirect contribution from the laser-induce orbital
magnetization
to the laser-induced torque is already contained in the response of
the torque operator $\vn{\mathcal{T}}$ to the applied laser field and
an extension of our formalism is not necessary. However, like in the
case of the SOT, it might become desirable in the future to develop
tools
to separate the laser-induced torque into spin and orbital
contributions.
We leave the development of such tools for future work.

\subsection{Wannier interpolation of SOI (WISOI)}
\label{sec_soi_matrix}
In the presence of SOI the electronic structure depends on the
direction of the (staggered) magnetization. When one would like
to evaluate Eq.~\eqref{eq_chi_noabrev} for many different 
magnetization directions one first needs to perform DFT
calculations of the electronic structure for all of them. If one
uses Wannier interpolation for computational speed-up of
Eq.~\eqref{eq_chi_noabrev} one additionally needs to compute 
MLWFs for all these magnetization directions. 
Alternatively, one may compute the electronic structure and MLWFs
without SOI and add the effect of SOI during the Wannier interpolation
of the material property tensors. This is a very convenient approach
when the material property tensors need to be evaluated for many
magnetization directions.

We briefly explain this Wannier interpolation of SOI here and
refer the reader to the appendix~\ref{sec_app_soi_matrix} for the
details of our implementation of this approach.
We denote the MLWFs without
SOI by $|\bar{W}_{\vn{R}n\sigma}\rangle$,
where $\sigma=\uparrow,\downarrow$ labels the spin.
The number of spin-up MLWFs is $N_{\uparrow}$
and the number of spin-down
MLWFs is $N_{\downarrow}$. The total number of
MLWFs is $N_{\rm W}=N_{\uparrow}+N_{\downarrow}$.
Using these MLWFs, we compute the matrix elements of 
the Hamiltonian $\bar{H}$ without SOI
\bege
\begin{aligned}
\bar{h}^{\sigma\sigma}_{\vn{R}nm}
&=\langle
\bar{W}_{\vn{0}n\sigma}
|\bar{H}|
\bar{W}_{\vn{R}m\sigma}
\rangle
\end{aligned}
\ee
and additionally the following 
matrix elements:
\bege\label{eq_prep_soi_mlwf}
\begin{aligned}
\mathcal{A}^{\sigma\sigma'j}_{\vn{R}nm}=
\sum_{\alpha}
\langle
\bar{W}_{\vn{0}n\sigma}
|\xi_{\alpha}(\vn{r}-\vn{r}_{\alpha}) L_{j}^{\alpha}|
\bar{W}_{\vn{R}m\sigma'}
\rangle,
\end{aligned}
\ee
where $L_{j}^{\alpha}$ is the orbital angular momentum operator of
atom $\alpha$ and $\xi_{\alpha}(\vn{r}-\vn{r}_{\alpha})$ is the SOI potential in a
sphere around atom $\alpha$ with its center at $\vn{r}_{\alpha}$.
The index $j$ takes the values $z$, $-$, and $+$, 
where $L^{\alpha}_{+}=(L^{\alpha}_{x}+iL^{\alpha}_{y})$
and $L^{\alpha}_{-}=(L^{\alpha}_{x}-iL^{\alpha}_{y})$.

One needs to add
$\mathcal{V}_{\rm SOI}=\sum_{\alpha}\xi_{\alpha}(\vn{r}-\vn{r}_{\alpha})\vn{L}^{\alpha}\cdot\vn{\sigma}$
to $\bar{H}$ in order to include SOI into the calculation.
To obtain the matrix 
elements $\mathcal{V}^{\sigma\sigma'}_{\vn{R}nm}=
\langle
\bar{W}_{\vn{0}n\sigma}
|
\mathcal{V}_{\rm SOI}
|
\bar{W}_{\vn{R}m\sigma'}
\rangle$
when the
N\'eel vector points into the 
direction $\hat{\vn{\mathcal{L}}}=(\sin\theta\cos\phi,\sin\theta\sin\phi,\cos\theta)^{\rm T}$
we need to multiply the Eq.~\eqref{eq_prep_soi_mlwf}
with trigonometric functions of the angles $\theta$ and $\phi$ as
follows:
\bege\label{eq_v_upup}
\begin{aligned}
\vn{\mathcal{V}}^{\uparrow\uparrow}_{\vn{R}}&=
\vn{\mathcal{A}}^{\uparrow\uparrow z}_{\vn{R}}\cos\theta+
\frac{1}{2}\vn{\mathcal{A}}^{\uparrow\uparrow -}_{\vn{R}}
\sin\theta \rme^{i\phi}+
\frac{1}{2}\vn{\mathcal{A}}^{\uparrow\uparrow +}_{\vn{R}}
\sin\theta \rme^{-i\phi},
\end{aligned}
\ee
\bege\label{eq_v_downdown}
\begin{aligned}
\vn{\mathcal{V}}^{\downarrow\downarrow}_{\vn{R}}&=
-\vn{\mathcal{A}}^{\downarrow\downarrow z}_{\vn{R}}\cos\theta-
\frac{1}{2}\vn{\mathcal{A}}^{\downarrow\downarrow -}_{\vn{R}}
\sin\theta \rme^{i\phi}-
\frac{1}{2}\vn{\mathcal{A}}^{\downarrow\downarrow +}_{\vn{R}}
\sin\theta \rme^{-i\phi},
\end{aligned}
\ee
\bege\label{eq_v_updown}
\begin{aligned}
\vn{\mathcal{V}}^{\uparrow\downarrow}_{\vn{R}}&=
-\vn{\mathcal{A}}^{\uparrow\downarrow z}_{\vn{R}}\sin\theta+
\vn{\mathcal{A}}^{\uparrow\downarrow -}_{\vn{R}}
\left[
\cos\frac{\theta}{2}
\right]^2 \rme^{i\phi}-
\vn{\mathcal{A}}^{\uparrow\downarrow +}_{\vn{R}}
\left[
\sin\frac{\theta}{2}
\right]^2 \rme^{-i\phi},
\end{aligned}
\ee
and
\bege\label{eq_v_downup}
\begin{aligned}
\vn{\mathcal{V}}^{\downarrow\uparrow}_{\vn{R}}&=
-\vn{\mathcal{A}}^{\downarrow\uparrow z}_{\vn{R}}\sin\theta-
\vn{\mathcal{A}}^{\downarrow\uparrow -}_{\vn{R}}
\left[\sin\frac{\theta}{2}
\right]^2 e^{i\phi}+
\vn{\mathcal{A}}^{\downarrow\uparrow +}_{\vn{R}}
\left[
\cos\frac{\theta}{2}
\right]^2 e^{-i\phi}.
\end{aligned}
\ee

Finally, the Wannier-interpolated Hamiltonian matrix including SOI is given by
\bege
H_{\vn{k}}^{\rm W}=
\begin{pmatrix}
\bar{\vn{h}}^{\uparrow\uparrow}_{\vn{k}} &0\\
0 &\bar{\vn{h}}^{\downarrow\downarrow}_{\vn{k}}
\end{pmatrix}+
\begin{pmatrix}
\vn{\mathcal{V}}^{\uparrow\uparrow}_{\vn{k}} 
&\vn{\mathcal{V}}^{\uparrow\downarrow}_{\vn{k}}\\
\vn{\mathcal{V}}^{\downarrow\uparrow}_{\vn{k}} 
&\vn{\mathcal{V}}^{\downarrow\downarrow}_{\vn{k}}
\end{pmatrix}
\ee
where the $N_{\sigma}\times N_{\sigma'}$ 
matrices $\bar{\vn{h}}^{\sigma\sigma'}_{\vn{k}}$  
and $\vn{\mathcal{V}}^{\sigma\sigma'}_{\vn{k}}$ are
the following Fourier transforms: 
\bege
\bar{\vn{h}}^{\sigma\sigma'}_{\vn{k}}=\sum_{\vn{R}}
\rme^{i\vn{k}\cdot\vn{R}}
\bar{\vn{h}}^{\sigma\sigma'}_{\vn{R}},\quad
\vn{\mathcal{V}}^{\sigma\sigma'}_{\vn{k}}=\sum_{\vn{R}}
\rme^{i\vn{k}\cdot\vn{R}}
\vn{\mathcal{V}}^{\sigma\sigma'}_{\vn{R}}.
\ee
With this Wannier-interpolated Hamiltonian $H_{\vn{k}}^{\rm W}$ we 
proceed in the usual way, i.e., it is diagonalized and the eigenvalues and
eigenvectors are used to evaluate Eq.~\eqref{eq_chi_noabrev}.
The velocity operator is obtained from $H_{\vn{k}}^{\rm W}$ as 
usual~\cite{wannier90communitycode}:
\bege
\vn{v}_{\vn{k}}=
\frac{1}{\hbar}
\frac{\partial
H_{\vn{k}}^{\rm W}
}
{\partial \vn{k}}.
\ee

\subsection{Symmetry}
\label{sec_symmetry}
In this section we discuss which components of the tensor
$\chi_{ijk}$, Eq.~\eqref{eq_chi_noabrev}, are allowed 
by symmetry in the Mn$_2$Au crystal. 
For this purpose we expand the laser-induced torque in orders of the staggered
magnetization $\mathcal{L}$ as follows:
\bege\label{eq_expansion_staggered}
T_i=\chi^{(\rm {4sp})}_{ijkl}E_j E^{*}_k \mathcal{L}_l
+\chi^{(\rm 5a)}_{ijklm}E_j E_k^{*} \mathcal{L}_l \mathcal{L}_m+\dots,
\ee 
 which implies that $\chi_{ijk}$ has the expansion 
\bege\label{eq_expand_chi_ijk}
\chi_{ijk}=\chi^{(\rm {4sp})}_{ijkl} \mathcal{L}_l
+\chi^{(\rm 5a)}_{ijklm} \mathcal{L}_l \mathcal{L}_m+\dots.
\ee

First we need to find out how the tensors $\chi^{(\rm {4sp})}_{ijkl}$
and $\chi^{(\rm 5a)}_{ijklm}$ transform under symmetry operations.
We recall that polar tensors satisfy
\bege
\chi^{\rm p}_{i'j'k'l'\dots}
=
\chi^{\rm p}_{ijkl\dots}
\mathcal{R}_{ii'}
\mathcal{R}_{jj'}
\mathcal{R}_{kk'}
\mathcal{R}_{ll'}
\dots
\ee
for all symmetry operations $\mathcal{R}$ in the point group of the
space group of the crystal, while
axial tensors satisfy
\bege
\chi^{\rm a}_{i'j'k'l'\dots}
=
(-1)^{\rm det \mathcal{R} }
\chi^{\rm a}_{ijkl\dots}
\mathcal{R}_{ii'}
\mathcal{R}_{jj'}
\mathcal{R}_{kk'}
\mathcal{R}_{ll'}
\dots.
\ee

In general, a staggered field, i.e., a field that
switches sign between different magnetic sublattices, transforms
differently from a non-staggered field. For this reason the
introduction
of polar and axial response tensors is insufficient to describe all
possible responses in antiferromagnets.
Therefore, we need to introduce two more types of tensors,
namely staggered polar and staggered axial tensors. 
In antiferromagnets with two magnetic sublattices that are not
related by a lattice translation these two additional types of tensors
may be defined as follows:
We call a tensor \textit{staggered polar}
when it satisfies
\bege
\chi^{\rm sp}_{i'j'k'l'\dots}
=
(-1)^{S(\mathcal{R})}
\chi^{\rm sp}_{ijkl\dots}
\mathcal{R}_{ii'}
\mathcal{R}_{jj'}
\mathcal{R}_{kk'}
\mathcal{R}_{ll'}
\dots
\ee
for all $\mathcal{R}$ in the point group.
Here, $S(\mathcal{R})=-1$ if $\mathcal{R}$
interchanges the two sublattices and
$S(\mathcal{R})=1$ otherwise.
Similarly, we call a tensor \textit{staggered axial}
when it satisfies
\bege
\chi^{\rm sa}_{i'j'k'l'\dots}
=
(-1)^{S(\mathcal{R})}(-1)^{\rm det \mathcal{R} }
\chi^{\rm sa}_{ijkl\dots}
\mathcal{R}_{ii'}
\mathcal{R}_{jj'}
\mathcal{R}_{kk'}
\mathcal{R}_{ll'}
\dots
\ee
for all $\mathcal{R}$ in the point group.

The N\'eel vector $\mathcal{L}$ transforms like a staggered axial tensor of rank 1.
Consequently, $\chi^{(\rm {4sp})}_{ijkl}$
is a staggered polar tensor of rank 4 and
$\chi^{(\rm 5a)}_{ijklm}$
is an axial tensor of rank 5.
Our expansion in Eq.~\eqref{eq_expansion_staggered} has the advantage
that it can be applied for a general direction of $\vn{\mathcal{L}}$
and that only information on the point group and on the positions of
the
magnetic sublattice sites are required to determine the tensors
allowed by
symmetry. In contrast, the usual symmetry analysis based on the magnetic point
groups suffers from the fact that in the presence of SOI the magnetic
point group depends on the direction of
$\vn{\mathcal{L}}$. Determining the response tensors allowed by the magnetic
point groups for selected high-symmetry directions of
$\vn{\mathcal{L}}$
provides less information than our expansion Eq.~\eqref{eq_expansion_staggered}.

Next, we need to find out which components of the tensors are
consistent with all symmetry operations $\mathcal{R}$. We find
that
in Mn$_2$Au staggered polar tensors of rank 4 are not allowed
by symmetry. However, axial tensors of rank 5 are allowed and we find
30 such tensors.
In order to discuss these tensors we introduce the notation
\bege\label{eq_notation_tensors}
\delta^{(ijklm)}_{nopqr}=\delta_{in}\delta_{jo}\delta_{kp}\delta_{lq}\delta_{mr}\rightarrow
\langle ijklm \rangle.
\ee
The 30 axial tensors of rank 5 allowed by symmetry in Mn$_2$Au are
listed
in table~\ref{tab_tensors}.

\begin{threeparttable}
\caption{
List of axial tensors of rank 5 allowed by symmetry in Mn$_2$Au.
The notation introduced in Eq.~\eqref{eq_notation_tensors} is used.
}
\label{tab_tensors}
\begin{ruledtabular}
\begin{tabular}{c|c|c||c|c|c|}
\#
&$\chi^{(\rm 5a)}_{ijklm}$&
&\#
&$\chi^{(\rm 5a)}_{ijklm}$&
\\
\hline
1 & $\langle13211\rangle-\langle23122\rangle$ &$\emptyset$
&16 &$\langle31233\rangle-\langle32133\rangle$ &$\emptyset$\\
\hline
2 &$\langle11213\rangle-\langle22123\rangle$ &$\nnearrow$
&17 &$\langle13233\rangle-\langle23133\rangle$ &$\perp$\\
\hline
3 & $\langle23212\rangle-\langle13121\rangle$ &$\parallel$
&18 &$\langle12333\rangle-\langle21333\rangle$ &\{17\}\\
\hline
4 &$\langle32221\rangle-\langle31112\rangle$ &$\parallel$
&19 &$\langle22231\rangle-\langle11132\rangle$ &\{6\}\\
\hline
5 &$\langle33213\rangle-\langle33123\rangle$ &$\nnearrow$
&20 &$\langle11231\rangle-\langle22132\rangle$ &\{2\}\\
\hline
6 & $\langle11123\rangle-\langle22213\rangle$ &$\nnearrow$
&21 &$\langle12232\rangle-\langle21131\rangle$ &$\nnearrow$\\
\hline
7 &$\langle12113\rangle-\langle21223\rangle$ &\{2\}
&22 &$\langle12223\rangle-\langle21113\rangle$ &\{21\}\\
\hline
8 & $\langle23221\rangle-\langle13112\rangle$ &\{3\}
&23 &$\langle11312\rangle-\langle22321\rangle$&\{3\}\\
\hline
9 &$\langle21311\rangle-\langle12322\rangle$ &$\parallel$
&24 &$\langle31211\rangle-\langle32122\rangle$& $\parallel$\\
\hline
10 &$\langle33231\rangle-\langle33132\rangle$ &\{5\}
&25 &$\langle33321\rangle-\langle33312\rangle$ &$\emptyset$\\
\hline
11 & $\langle12131\rangle-\langle21232\rangle$ &\{2\}
&26 &$\langle13222\rangle-\langle23111\rangle$&\{9\}\\
\hline
12 &$\langle31332\rangle-\langle32331\rangle$ &\{5\}
&27 &$\langle32212\rangle-\langle31121\rangle$ &\{4\}\\
\hline
13 & $\langle13332\rangle-\langle23331\rangle$ &$\nnearrow$
&28 &$\langle12311\rangle-\langle21322\rangle$&$\emptyset$\\
\hline
14 &$\langle31323\rangle-\langle32313\rangle$ &\{5\}
&29 &$\langle11321\rangle-\langle22312\rangle$ &\{3\}\\
\hline
15 &$\langle13323\rangle-\langle23313\rangle$ &\{13\}
&30 &$\langle31222\rangle-\langle32111\rangle$ &\{24\}\\
\end{tabular}
\end{ruledtabular}
\end{threeparttable}

Since only torques that are perpendicular to the N\'eel vector
are relevant, we do not need to consider tensor components,
where the first index is equal to the last two indices. Therefore,
we may ignore the tensors 1, 16, and 28 (indicated by $\emptyset$ in
the table). Additionally, we may ignore tensor 25, because the
indices $l$ and $m$ are interchangeable in the second term on the
right-hand side of Eq.~\eqref{eq_expand_chi_ijk}.
The number of tensors that need to be considered may be reduced
further
by noting that the two indices $l$ and $m$ in $\chi^{(\rm
  5a)}_{ijklm}$
are both contracted with the staggered magnetization,
while the indices $j$ and $k$ are both contracted with the
electric field. Therefore, when the tensors in Table~\ref{tab_tensors}
are inserted into Eq.~\eqref{eq_expand_chi_ijk} several of them
are effectively equivalent. The tensors that do not need to be
considered
due to this are denoted by $\{r\}$ in the Table, where $r$ is the
number of the tensor that can be used to replace it.

In Mn$_{2}$Au the N\'eel vector $\vn{\mathcal{L}}$ lies in the $xy$ plane. 
Therefore, we
first
discuss possible tensors $\chi^{(\rm 5a)}_{ijklm}$ where the indices 4
and 5 take values corresponding to in-plane $\vn{\mathcal{L}}$.
These are the tensors 3, 4, 9, 24
(indicated by $\parallel$ in the table).
Tensor 4 predicts torques
in the $z$ direction for light polarized linearly in the $x$ or in the
$y$ direction. 
Note that the torque from tensor 4 vanishes 
when the in-plane N\'eel vector is
parallel or perpendicular to the $x$ axis, i.e., when either $\mathcal{L}_{x}=0$
or $\mathcal{L}_{y}=0$.

Of course, the tensor 4 predicts torques also for
circularly polarized light. For example, tensor 4 predicts torques
for light circularly polarized in the $yz$ plane or in $xz$ plane.
However, the response is even in the light
helicity $\lambda$. 
Note that for light circularly polarized in the $xy$ plane the two terms in
tensor
4 -- $\langle 32221 \rangle$ and $\langle 31112 \rangle$ -- cancel
each other.

Tensor 3 predicts a torque in $y$ direction for linearly polarized
light
with $\vn{\epsilon}$ in the $yz$ plane and a torque in $x$ direction
for linearly polarized
light
with $\vn{\epsilon}$ in the $zx$ plane.
Tensor 9 predicts a torque in the $y$ direction
when the magnetization is along the $x$ direction and when $\vn{\epsilon}$
lies in the $zx$ plane and it predicts a torque in the $x$
direction when the magnetization is along the $y$ direction and when $\vn{\epsilon}$
lies in the $yz$ plane.

While the N\'eel vector in Mn$_2$Au lies in the $xy$ plane, there might be AFMs with the
same crystal structure as Mn$_2$Au but with a different magnetic
anisotropy, for which $\mathcal{L}_z\ne 0$ might be relevant.
Therefore, we discuss the cases with $\mathcal{L}_z\ne 0$ in the following.
The case with $\hat{\vn{\mathcal{L}}}\Vert \hat{\vn{e}}_{z}$ is described by
the tensor 17 (indicated by $\perp$ in the table).
This tensor predicts a torque in the $x$ direction when
$\vn{\epsilon}$ lies in the $yz$ plane and it predicts a torque in $y$
direction
when $\vn{\epsilon}$ lies in the $zx$ plane.

\section{Results}
\label{sec_results}

\subsection{Computational details}
\label{sec_computational_parameters}
We employ the full-potential linearized augmented-plane-wave (FLAPW)
program {\tt FLEUR}~\cite{fleurcode} in order to determine the
electronic structure of Mn$_2$Au selfconsistently
within the generalized-gradient
approximation~\cite{PerdewBurkeEnzerhof} to
density-functional theory.
The space group of Mn$_2$Au is I4/mmm.
We use the experimental lattice parameters, which are
$a=3.328$~\AA\, and $c=8.539$~\AA\, 
for the conventional tetragonal
unit cell containing 2 formula units~\cite{Wells:a07238}.
In our calculations we use the primitive unit cell,
which contains 2 Mn atoms.
We perform calculations with and without SOI in order to
test the accuracy of the WISOI.
The magnetic moments of the Mn atoms are 3.73$\mu_{\rm B}$.

In order to perform the Brillouin zone integrations
in Eq.~\eqref{eq_chi_noabrev} 
computationally efficiently
based on the Wannier interpolation technique,
we constructed 18 MLWFs per
transition metal
atom from an $8\times 8\times 8$
$\vn{k}$~mesh~\cite{wannier90communitycode,WannierPaper}
when SOI is included in the calculations.
In order to prepare MLWFs for the WISOI approach we constructed
9 MLWFs per transition metal atom and per spin. While these are in
total
18 MLWFs per transition metal atom as well, two separate runs of
Wannier90 are performed, one for spin-up and one for spin-down.
The numerical effort needed to obtain two sets of MLWFs without SOI, where
each set contains 9 MLWFs per transition metal
atom
and per spin, is smaller than the numerical effort needed to obtain 18
MLWFs with SOI per transition metal atom, because the spin-up and
spin-down
bands are decoupled in the former case.

\subsection{Laser-induced torques at optical frequencies}
\label{sec_laser_induced_torques}

\begin{figure}
  \includegraphics[height=0.8\linewidth]{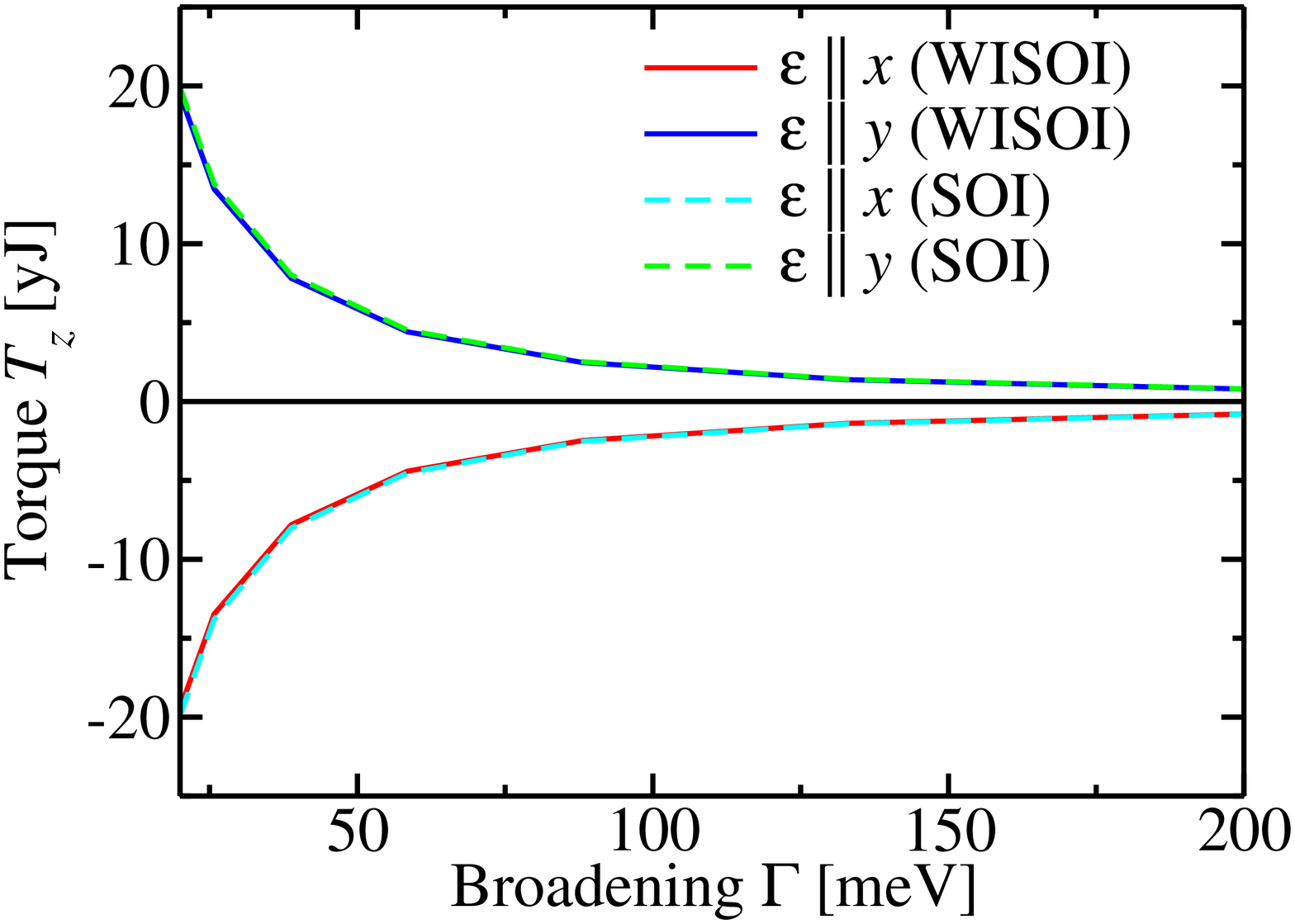}
\caption{\label{fig1}
  Laser-induced torque in Mn$_2$Au
  vs.\ quasi-particle broadening $\Gamma$
when the N\'eel vector points in the 110 direction.
Comparison between the results obtained within
the WISOI approach (solid lines) and the results
obtained based on MLWFs that include SOI (dashed lines).
The torque is shown in units of yJ=yoctojoule=10$^{-24}$Joule. 
}
\end{figure}

In this section we discuss laser induced torques for the laser
intensity $I=10$~GW/cm$^{2}$
when the photon energy is set to 1.55~eV.
In Fig.~\ref{fig1} we show the
laser-induced
torque as a function of $\Gamma$ when the N\'eel vector points
in the 110 direction. This torque is consistent with
tensor 4 in Table~\ref{tab_tensors}, which  describes a laser-induced torque in
the $z$ direction that differs in sign for linearly polarized light
along
the $x$ and $y$ directions.
The figure shows both the results obtained within
the WISOI approach (solid lines) and the results
obtained based on MLWFs that include SOI (dashed lines). The very good
agreement
between the two approaches proves the validity and accuracy of the
WISOI approach. Therefore, all following figures below show only the WISOI results.
In Fig.~\ref{fig1} we present the torque in units of yoctojoule. The
effective staggered
magnetic field that produces a torque of one yoctojoule is 14.5~mT.
Thus, the torques shown in Fig.~\ref{fig1} are of the same order of
magnitude as the laser-induced torques in the ferromagnets
Fe, Co, and FePt that we studied in Ref.~\cite{lasintor}.

Tensor 4 predicts a torque in the $z$ direction also for light
circularly
polarized in the $yz$ or $zx$ planes. Our calculations confirm this
prediction. For light circularly polarized in the $zx$ plane we find
$T_z<0$ and its magnitude is half of the magnitude for linearly
polarized light with $\vn{\epsilon}\Vert\hat{\vn{e}}_{x}$.
Similarly, for light circularly polarized in the $yz$ plane we find
$T_z$ to be half of what it is when $\vn{\epsilon}\Vert\hat{\vn{e}}_{y}$.

\begin{figure}
  \includegraphics[height=0.8\linewidth]{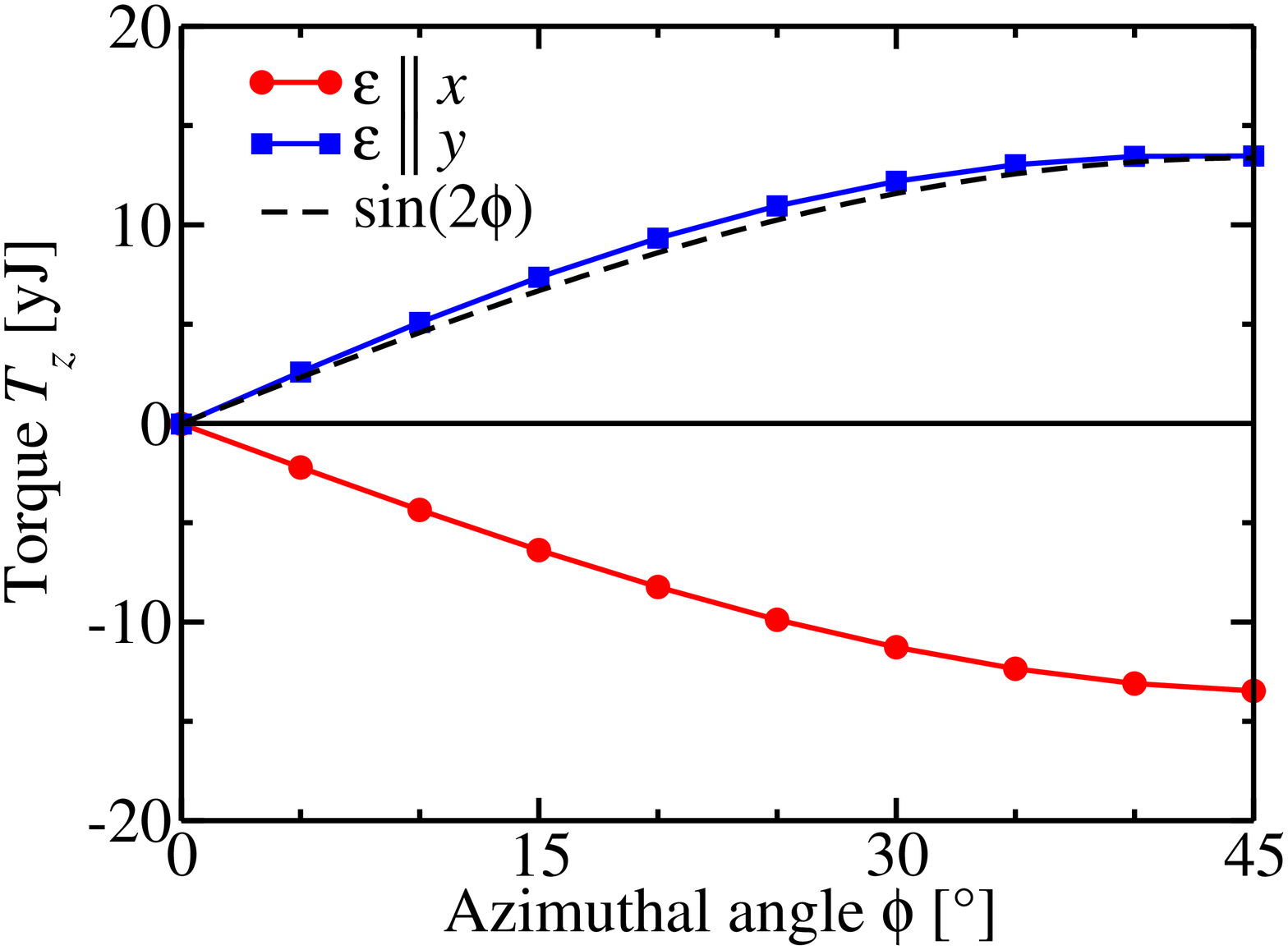}
\caption{\label{fig2}
  Laser-induced torque in Mn$_2$Au
  vs.\ azimuthal angle $\phi$ when the polar angle is $\theta=90^{\circ}$ 
and when $\Gamma=25$~meV.
}
\end{figure}

In order to discuss the dependence of the laser-induced torque
on the N\'eel  vector we introduce the azimuthal angle $\phi$
and the polar angle $\theta$ such that $\hat{\vn{\mathcal{L}}}=(\sin\theta\cos\phi,\sin\theta\sin\phi,\cos\theta)^{\rm T}$.
In Fig.~\ref{fig2} we show the
laser-induced
torque as a function of the azimuthal angle $\phi$ when
$\theta=90^{\circ}$ 
and when the
quasiparticle broadening is set to $\Gamma=25$~meV.
Tensor 4 predicts the $\phi$ dependence
$\propto\hat{M}_{x}\hat{M}_{y}\propto \sin\phi \cos\phi\propto
\sin(2\phi)$,
which is illustrated in Fig.~\ref{fig2} by the dashed line and which
fits the \textit{ab-initio} data very well.

\begin{figure}
  \includegraphics[height=0.8\linewidth]{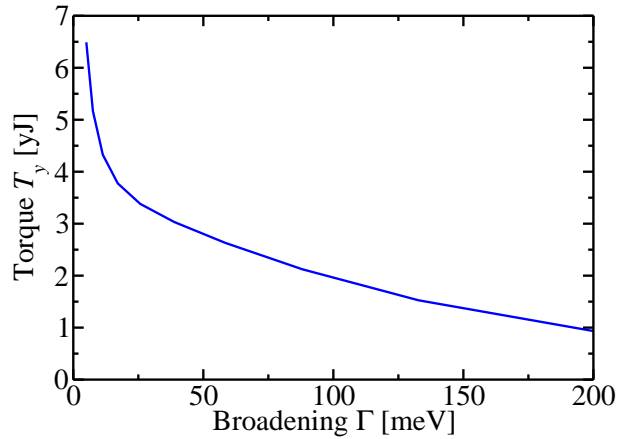}
\caption{\label{fig_torquey_magx}
  Laser-induced torque in Mn$_2$Au
  vs.\ quasi-particle broadening $\Gamma$
when the staggered magnetization is along (100), i.e.,
$\hat{\vn{\mathcal{L}}}=\hat{\vn{e}}_{x}$,
and when $\vn{\epsilon}=(\hat{\vn{e}}_{x}+\hat{\vn{e}}_{z})/\sqrt{2}$,
i.e., $\vn{\epsilon}\Vert(101)$.
}
\end{figure}

In Fig.~\ref{fig_torquey_magx} we plot the
component $T_{y}$ when the magnetization is along the $x$ direction 
and when $\vn{\epsilon}=(\hat{\vn{e}}_{x}+\hat{\vn{e}}_{z})/\sqrt{2}$
i.e., $\vn{\epsilon}\Vert(101)$. This torque is
consistent with tensor 9, which predicts a torque in $y$ direction
when the magnetization is along $x$ direction and when $\vn{\epsilon}$
lies in the $zx$ plane.

\begin{figure}
  \includegraphics[height=0.8\linewidth]{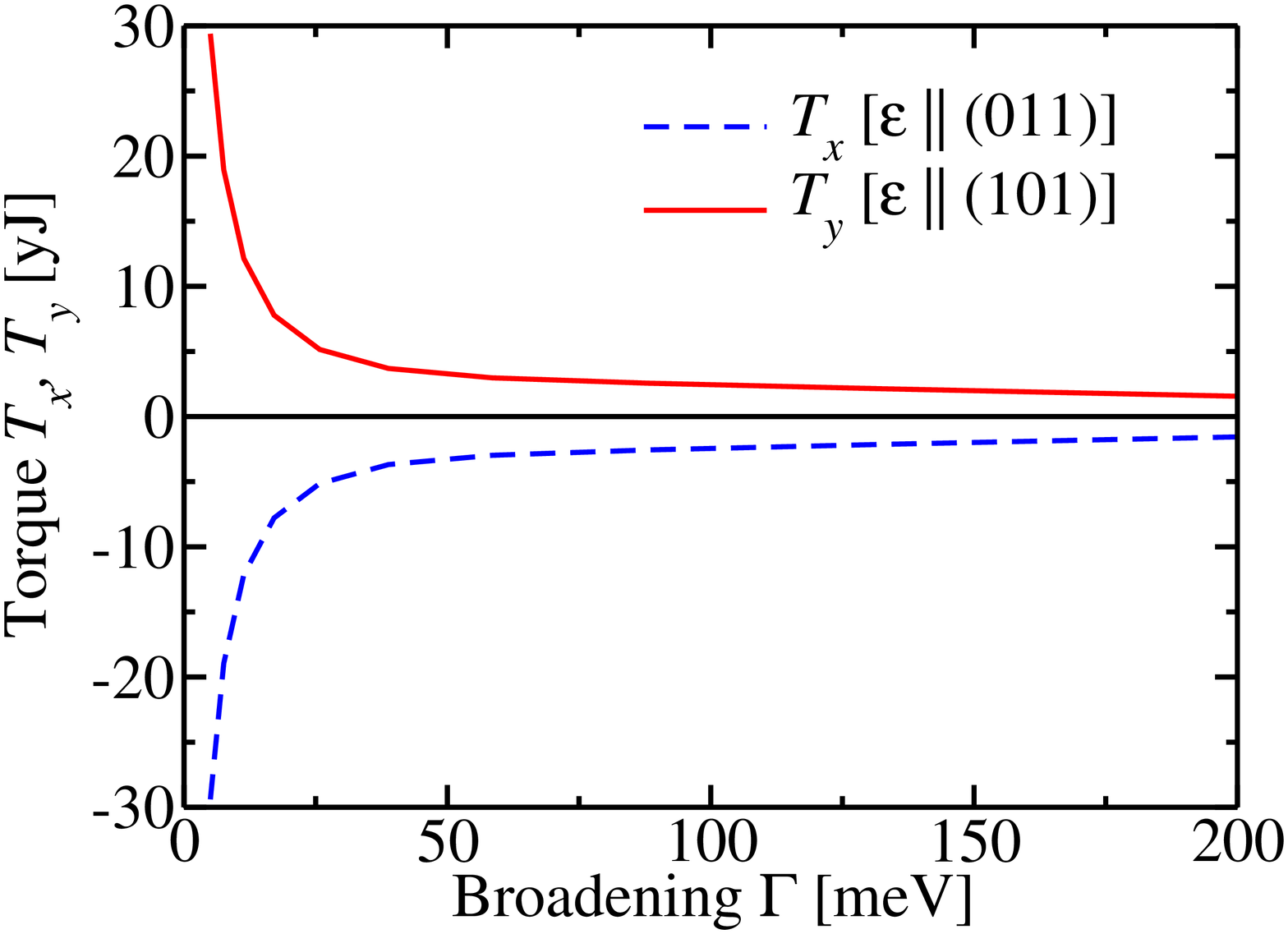}
\caption{\label{fig_neel_z}
  Laser-induced torque in Mn$_2$Au
  vs.\ quasi-particle broadening $\Gamma$
when the staggered magnetization is along (001), i.e., $\hat{\vn{\mathcal{L}}}=\hat{\vn{e}}_{z}$.
}
\end{figure}

Next, we discuss the case $\hat{\vn{\mathcal{L}}}=\hat{\vn{e}}_{z}$.
In Fig.~\ref{fig_neel_z} we show $T_{y}$  for
$\vn{\epsilon}=(\hat{\vn{e}}_{x}+\hat{\vn{e}}_{z})/\sqrt{2}$,
i.e., $\vn{\epsilon}\Vert(101)$,
and $T_{x}$  for
$\vn{\epsilon}=(\hat{\vn{e}}_{y}+\hat{\vn{e}}_{z})/\sqrt{2}$
i.e., $\vn{\epsilon}\Vert(011)$.
The torques $T_{x}$ and $T_{y}$ are equal but opposite in the
figure, consistent with the tensor 17 in
Table~\ref{tab_tensors}.

\subsection{Laser-induced torques at THz frequencies}
\label{sec_laser_induced_torques_thz}

\begin{figure}
  \includegraphics[height=0.8\linewidth]{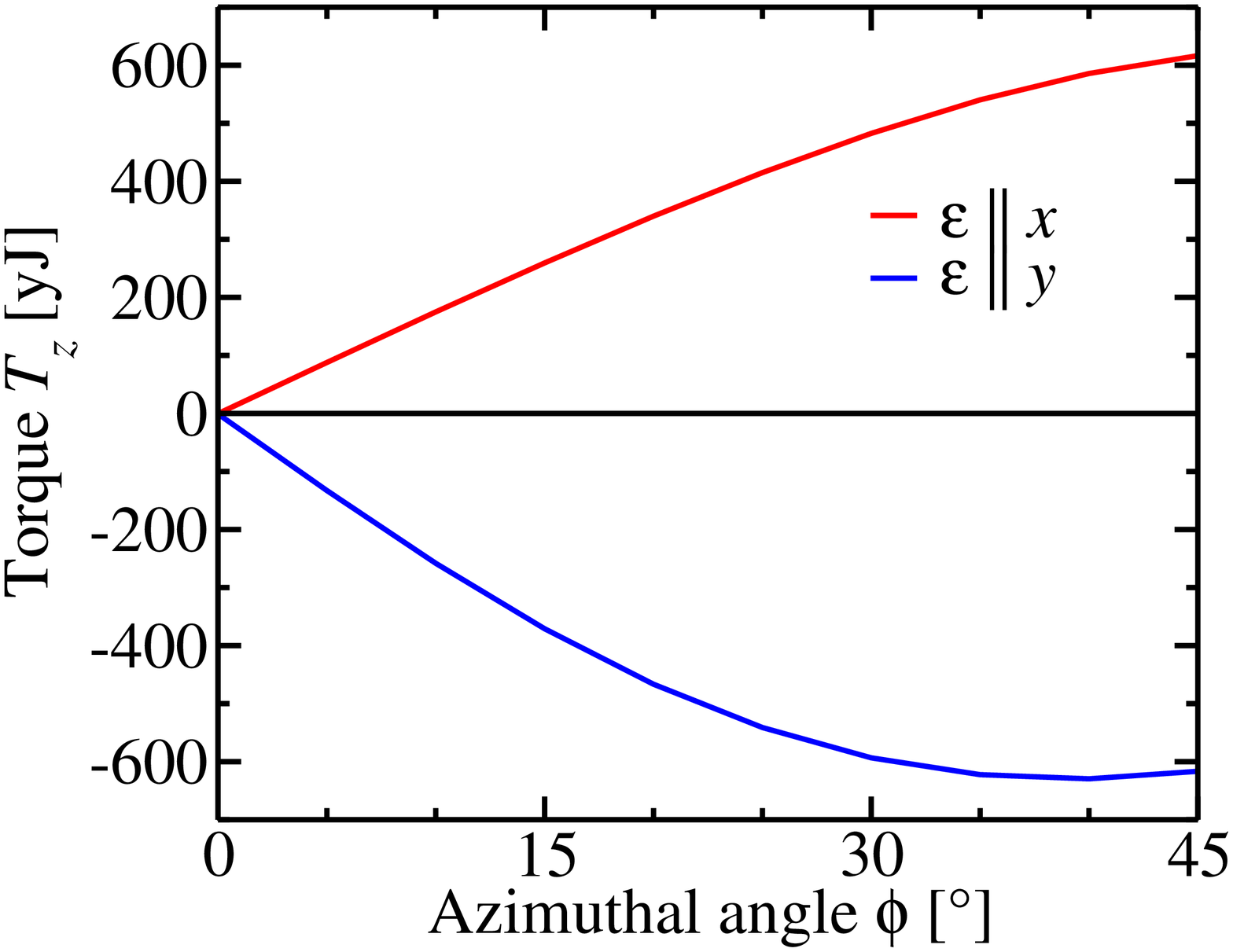}
\caption{\label{fig_thz_vs_angle}
  Laser-induced torque in Mn$_2$Au
  vs.\ azimuthal angle $\phi$ when the polar angle is $\theta=90^{\circ}$ 
and when $\Gamma=25$~meV.
}
\end{figure}

In Fig.~\ref{fig_thz_vs_angle} we show the laser-induced torque
for a THz-laser with frequency of 1~THz for the laser
intensity $I=10$~GW/cm$^{2}$.
The intensity $I=10$~GW/cm$^2$ corresponds to the electric
field of the laser light of $E=2.7$~MV/cm.
The SOT in Mn$_2$Au is described by the odd torkance $t^{\rm
  odd}=0.6ea_{0}$
at $\Gamma=25$~meV. The corresponding torque 
at $E=2.7$~MV/cm is 1370~yJ. This is larger than the maximum
laser-induced
torque in Fig.~\ref{fig_thz_vs_angle} by a factor of 2 only.
At higher broadening $\Gamma=130$~meV
we find a SOT of  230~yJ ($t^{\rm odd}=0.1ea_{0}$)
and a laser-induced torque of 12~yJ. The ratio of SOT to the laser-induced 
torque is thus increased to a factor of 20 by the increase of broadening.
At even higher broadening $\Gamma=200$~meV
we find a SOT of 132~yJ ($t^{\rm odd}=0.058ea_{0}$)
and a laser-induced torque of 2.6~yJ, i.e., the ratio of SOT to the
laser-induced torque is further increased to 50.

In the experiment on CuMnAs in Ref.~\cite{THz_writing_speed_afm}
it was shown that THz electric field pulses switch the N\'eel vector.
This switching was attributed to the action of the SOT.
We therefore compare the magnitudes of the SOT and of the
laser-induced torque in Mn$_2$Au in order to judge if the
laser-induced
torque might contribute as well to the switching by THz pulses.
In Ref.~\cite{THz_writing_speed_afm} the
maximum amplitude of the applied THz electric field is $E=0.11$~MV/cm,
which is smaller than $E=2.7$~MV/cm by a factor 24.5 and which 
corresponds to an intensity of 0.017~GW/cm$^2$ only. At such a small
intensity the ratio of the SOT to
the laser-induced torque is roughly 50 in Mn$_2$Au, such
that
the laser-induced torque is negligible in Mn$_2$Au already at
$\Gamma=25$~meV.
As discussed in the previous paragraph the ratio of the SOT to the
laser-induced torque will futher increase by an order of magnitude
when
the broadening is increased to  $\Gamma=130$~meV such that the
contribution of the laser-induced torques relative to the SOT
decreases
even further with increasing broadening.
Since the laser-induced torque
scales quadratically with the electric field of the laser, it becomes
more
important relative to the SOT at higher fields. However, picosecond THz-pulses
of intensities
as high as $I=10$~GW/cm$^2$ are far above the damage threshold of
CuMnAs.

While literature values for the damage threshold of Mn$_2$Au are not
available,
we assume that  picosecond THz-pulses
of intensities
as high as $I=10$~GW/cm$^2$ cannot be applied to Mn$_2$Au without
damaging it. In order to apply such high laser intensities without
damaging
the metallic AFMs much shorter femtosecond pulses are necessary.
Therefore, we expect the SOTs from THz laser pulses to be more
important than the laser-induced torques. However, when lasers with optical
frequencies are used,
the SOT oscillates with the frequency of the laser field, which is
much higher than the AFM resonance frequencies and therefore the SOT cannot
excite the AFM. Therefore, when lasers in the optical range are used
the laser-induced torques are important while the SOT is irrelevant.

\section{Summary}
\label{sec_summary}
We compute the laser-induced torques in the AFM Mn$_2$Au based on
the Keldysh nonequilibrium approach. We find the laser-induced torques
to be of the same order of magnitude as those in the ferromagnets Fe,
Co,
and FePt. From the phenomenological theory of the IFE in non-magnets
one intuitively expects that laser-induced torques can be generated only by
circularly polarized light. In contrast, we find that linearly
polarized light is sufficient to stimulate torques in Mn$_2$Au. 
We corroborate this finding
by a detailed symmetry analysis of
the laser-induced torque.
Additionally, we discuss the laser-induced torques at
THz frequencies. At THz frequencies we compare the laser-induced
torque
to the SOT and find the SOT to be larger than the laser-induced torque
by a factor of 50 at the light intensity used in experiments. In
contrast,
at optical frequencies only the laser-induced torque induces
magnetization
dynamics and the SOT may be neglected.
In order to compute response coefficients from Wannier interpolation
conveniently for many magnetization directions we develop the WISOI
approach. We demonstrate that the WISOI approach reproduces the
response coefficients obtained from MLWFs that include SOI with high
accuracy.

\section*{Acknowledgments}We acknowledge financial support from
Leibniz Collaborative Excellence project OptiSPIN $-$ Optical Control
of Nanoscale Spin Textures, and funding  under SPP 2137 ``Skyrmionics"
of the DFG. 
We gratefully acknowledge financial support from the European Research
Council (ERC) under 
the European Union's Horizon 2020 research and innovation 
program (Grant No. 856538, project ``3D MAGiC''). 
The work was also supported by the Deutsche Forschungsgemeinschaft 
(DFG, German Research Foundation) $-$ TRR 173 $-$ 268565370 (project
A11), TRR 288 $-$ 422213477 (project B06).  We  also gratefully
acknowledge the J\"ulich 
Supercomputing Centre and RWTH Aachen University for providing
computational 
resources under project No. jiff40.

\appendix

\section{Spin-orbit coupling matrix elements}
\label{sec_app_soi_matrix}

In this appendix we describe
a procedure that allows us to use MLWFs that are
calculated without SOI  in order to compute material property tensors
including the effect of SOI by adding the effect of SOI during the
Wannier interpolation. The advantage of this procedure is that
only two sets of MLWFs need to be calculated: One set
of MLWFs for the spin-up states and
a second set of MLWFs for the spin-down states. 

We decompose the Hamiltonian as follows:
\bege
H=\bar{H}+\mathcal{V}_{\rm SOI},
\ee
where $\mathcal{V}_{\rm SOI}$ denotes SOI and
$\bar{H}$ is the Hamiltonian without SOI.
In order to obtain MLWFs without 
SOI, $\mathcal{V}_{\rm SOI}$ is ignored during the 
generation of MLWFs.
We denote the MLWFs without
SOI by $|\bar{W}_{\vn{R}n\sigma}\rangle$, 
where $\sigma=\uparrow,\downarrow$ labels the spin.
The number of spin-up MLWFs is $N_{\uparrow}$
and the number of spin-down 
MLWFs is $N_{\downarrow}$. The total number of
MLWFs is $N_{\rm W}=N_{\uparrow}+N_{\downarrow}$.
The MLWFs are related to the Bloch functions by the
transformation~\cite{wannier90communitycode}
\bege
|\bar{W}_{\vn{R}n\sigma}\rangle=\frac{1}{\mathcal{N}}
\sum_{\vn{k}}
       e^{-i\vn{k}\cdot\vn{R}}
       \sum_{m}U_{mn\sigma}^{(\vn{k})}
       |\bar{\psi}_{\vn{k}m\sigma}\rangle,
\ee
where $\mathcal{N}$ is the number of $\vn{k}$-points used
for the generation of the MLWFs and
where the Bloch states $|\bar{\psi}_{\vn{k}m\sigma}\rangle$
are the eigenstates of the Hamiltonian $\bar{H}$ without SOI
with corresponding eigenvalues $\bar{\mathcal{E}}_{\vn{k}n\sigma}$,
i.e.,
\bege
\bar{H}|\bar{\psi}_{\vn{k}n\sigma}\rangle=
\bar{\mathcal{E}}_{\vn{k}n\sigma}
|\bar{\psi}_{\vn{k}n\sigma}\rangle.
\ee
The Wannier90 code~\cite{wannier90communitycode} 
determines the 
transformation $U_{mn\sigma}^{(\vn{k})}$ such that the
resulting Wannier functions are maximally localized.

The Wannier-interpolated Hamiltonian 
matrix $\bar{\vn{H}}^{\rm W}_{\vn{k}}$
without SOI is
given by the Fourier transform~\cite{wannier90communitycode}
\bege
\bar{\vn{H}}^{\rm W}_{\vn{k}}=
\begin{pmatrix}
\bar{\vn{h}}^{\uparrow\uparrow}_{\vn{k}} &0 \\
0 &\bar{\vn{h}}^{\downarrow\downarrow}_{\vn{k}}
\end{pmatrix}=
\sum_{\vn{R}}\rme^{i\vn{k}\cdot\vn{R}}
\begin{pmatrix}
\bar{\vn{h}}^{\uparrow\uparrow}_{\vn{R}} &0\\
0 &\bar{\vn{h}}^{\downarrow\downarrow}_{\vn{R}}
\end{pmatrix},
\ee
where $\bar{\vn{h}}^{\uparrow\uparrow}_{\vn{k}}$ 
and $\bar{\vn{h}}^{\uparrow\uparrow}_{\vn{R}}$
are $N_{\uparrow}\times N_{\uparrow}$-matrices 
and $\bar{\vn{h}}^{\downarrow\downarrow}_{\vn{k}}$ 
and $\bar{\vn{h}}^{\downarrow\downarrow}_{\vn{R}}$
are $N_{\downarrow}\times N_{\downarrow}$-matrices.
$\bar{\vn{H}}_{\vn{k}}^{\rm W}$ is a $N_{\rm W}\times N_{\rm W}$-matrix.
The matrix elements of $\bar{\vn{h}}^{\sigma\sigma}_{\vn{R}}$
are given by~\cite{wannier90communitycode}
\bege
\begin{aligned}
\bar{h}^{\sigma\sigma}_{\vn{R}nm}
&=\langle
\bar{W}_{\vn{0}n\sigma}
|\bar{H}|
\bar{W}_{\vn{R}m\sigma}
\rangle=\\
&=\frac{1}{\mathcal{N}}\sum_{\vn{k}m'}
\bar{\mathcal{E}}_{\vn{k}m'\sigma}e^{-i\vn{k}
\cdot \vn{R}}
\left(U^{(\vn{k})}_{m'n\sigma}\right)^{*}
U_{m'm\sigma}^{(\vn{k})}.
\end{aligned}
\ee

%SOI is given by 
%\bege
%\xi
%[
%L_{z}\sigma_{z}
%+
%\frac{1}{2}
%L_{+}\sigma_{-}
%+
%\frac{1}{2}
%L_{-}\sigma_{+}
%]
%\ee

Since SOI is important only close to the
nuclei, we may neglect 
SOI in the interstitial regions between the atoms.
We may therefore express SOI as
\bege\label{eq_soi_ham}
\mathcal{V}_{\rm SOI}=\sum_{\alpha}\xi_{\alpha}(\vn{r}-\vn{r}_{\alpha})\vn{L}^{\alpha}
\cdot\vn{\sigma},
\ee
where 
\bege
\xi_{\alpha}(\vn{r})=
\left\{
\begin{matrix}
\frac{\mu_{\rm B}}{2c r}\frac{dV_{\alpha}(r)}{dr}   {\rm \,\, for\,\,} |\vn{r}|<R_{\alpha} \\
0 {\rm \,\, for\,\,} |\vn{r}|>R_{\alpha}
\end{matrix}
\right.
\ee
is the SOI potential in a sphere around atom $\alpha$,
$R_{\alpha}$ is the atomic radius of atom $\alpha$,
$\vn{r}_{\alpha}$ is the position of the center of atom $\alpha$,
$\vn{L}^{\alpha}=(\vn{r}-\vn{r}_{\alpha})\times\vn{p}$ 
is the angular momentum operator
of atom $\alpha$, and $V_{\alpha}(r)$ is the spherical part of the
potential in the sphere around atom $\alpha$.

In order to add the effect of SOI the following matrix elements need to
be evaluated in the basis of Bloch functions:
\bege
\mathcal{A}^{\sigma\sigma'-}_{\vn{k}nm}=
\sum_{\alpha}
\langle
\bar{\psi}_{\vn{k}n\sigma}
|\xi_{\alpha}(\vn{r}-\vn{r}_{\alpha}) L^{\alpha}_{-}|
\bar{\psi}_{\vn{k}m\sigma'}
\rangle,
\ee
\bege
\mathcal{A}^{\sigma\sigma'+}_{\vn{k}nm}=
\sum_{\alpha}
\langle
\bar{\psi}_{\vn{k}n\sigma}
|\xi_{\alpha}(\vn{r}-\vn{r}_{\alpha}) L_{+}^{\alpha}|
\bar{\psi}_{\vn{k}m\sigma'}
\rangle,
\ee
and
\bege
\mathcal{A}^{\sigma\sigma'z}_{\vn{k}nm}=
\sum_{\alpha}
\langle
\bar{\psi}_{\vn{k}n\sigma}
|\xi_{\alpha}(\vn{r}-\vn{r}_{\alpha}) L_{z}^{\alpha}|
\bar{\psi}_{\vn{k}m\sigma'}
\rangle,
\ee
where $L^{\alpha}_{+}=(L^{\alpha}_{x}+iL^{\alpha}_{y})$ 
and $L^{\alpha}_{-}=(L^{\alpha}_{x}-iL^{\alpha}_{y})$.
These matrix elements satisfy the relations
\bege
\begin{aligned}
&\mathcal{A}^{\uparrow\uparrow z}_{\vn{k}mn}=
\left[\mathcal{A}^{\uparrow\uparrow z}_{\vn{k}nm}\right]^{*},
&\mathcal{A}^{\uparrow\uparrow +}_{\vn{k}mn}=
\left[
\mathcal{A}^{\uparrow\uparrow -}_{\vn{k}nm}
\right]^{*},\\
&\mathcal{A}^{\downarrow\downarrow z}_{\vn{k}mn}=
\left[\mathcal{A}^{\downarrow\downarrow z}_{\vn{k}nm}\right]^{*},
&\mathcal{A}^{\downarrow\downarrow +}_{\vn{k}mn}=
\left[
\mathcal{A}^{\downarrow\downarrow -}_{\vn{k}nm}
\right]^{*}
,\\
&\mathcal{A}^{\downarrow\uparrow z}_{\vn{k}mn}=
\left[\mathcal{A}^{\uparrow\downarrow z}_{\vn{k}nm}\right]^{*},
&\mathcal{A}^{\downarrow\uparrow +}_{\vn{k}mn}=
\left[
\mathcal{A}^{\uparrow\downarrow -}_{\vn{k}nm}
\right]^{*}.\\
\end{aligned}
\ee
Subsequently, these matrix elements are transformed into the
MLWF basis:
\bege\label{eq_app_ksi_R}
\begin{aligned}
\mathcal{A}^{\sigma\sigma'j}_{\vn{R}nm}=
\sum_{\alpha}
\langle
\bar{W}_{\vn{0}n\sigma}
|\xi_{\alpha}(\vn{r}-\vn{r}_{\alpha}) L_{j}^{\alpha}|
\bar{W}_{\vn{R}m\sigma'}
\rangle
\\=
\frac{1}{\mathcal{N}}
\sum_{\vn{k}n'm'}e^{-i\vn{k}\cdot\vn{R}}
[U^{(\vn{k})}_{n'n\sigma}]^{*}
U^{(\vn{k})}_{m'm\sigma'}
\mathcal{A}^{\sigma\sigma'j}_{\vn{k}n'm'},
\end{aligned}
\ee
where $j=-,z,+$.

When the (staggered) magnetization points into the 
direction $\hat{\vn{\mathcal{L}}}=(\sin\theta\cos\phi,\sin\theta\sin\phi,\cos\theta)^{\rm
  T}$ the spinor of the spin-up electrons is
\bege
|\uparrow,\theta,\phi\rangle=
\left(
\begin{array}{c}
e^{-i\phi/2}\cos(\theta/2)\\
e^{i\phi/2}\sin(\theta/2)
\end{array}
\right)
\ee
while the spinor of the spin-down electrons is
\bege
|\downarrow,\theta,\phi\rangle=
\left(
\begin{array}{c}
-e^{-i\phi/2}\sin(\theta/2)\\
e^{i\phi/2}\cos(\theta/2)
\end{array}
\right).
\ee
We define
\bege
\begin{aligned}
\sigma_{+}=\sigma_x+i\sigma_y, \quad \sigma_{-}=\sigma_x-i\sigma_y.
\end{aligned}
\ee
The matrix elements of $\sigma_z$ are
\bege\label{eq_sigma_z}
\begin{aligned}
&\langle
\uparrow,\theta,\phi
|
\sigma_{z}
|
\uparrow,\theta,\phi
\rangle=\cos\theta,\\
&\langle
\downarrow,\theta,\phi
|
\sigma_{z}
|
\downarrow,\theta,\phi
\rangle=-\cos\theta,\\
&\langle
\uparrow,\theta,\phi
|
\sigma_{z}
|
\downarrow,\theta,\phi
\rangle
%=-2
%\sin\frac{\theta}{2}
%\cos\frac{\theta}{2}
=-\sin\theta,\\
&\langle
\downarrow,\theta,\phi
|
\sigma_{z}
|
\uparrow,\theta,\phi
\rangle=-\sin\theta.
\end{aligned}
\ee
The matrix elements of $\sigma_{+}$ are
\bege\label{eq_sigma_plus}
\begin{aligned}
&\langle
\uparrow,\theta,\phi
|
\sigma_{+}
|
\uparrow,\theta,\phi
\rangle=
%\frac{1}{2}
\sin\theta\rme^{i\phi},\\
&\langle
\downarrow,\theta,\phi
|
\sigma_{+}
|
\downarrow,\theta,\phi
\rangle=-\sin\theta\rme^{i\phi},\\
&\langle
\downarrow,\theta,\phi
|
\sigma_{+}
|
\uparrow,\theta,\phi
\rangle=-2
\left[
\sin\frac{\theta}{2}
\right]^2
\rme^{i\phi},\\
&\langle
\uparrow,\theta,\phi
|
\sigma_{+}
|
\downarrow,\theta,\phi
\rangle=2\left[
\cos\frac{\theta}{2}
\right]^2
\rme^{i\phi}.
\end{aligned}
\ee
The matrix elements of $\sigma_{-}$ are
\bege\label{eq_sigma_minus}
\begin{aligned}
&\langle
\uparrow,\theta,\phi
|
\sigma_{-}
|
\uparrow,\theta,\phi
\rangle=\sin\theta\rme^{-i\phi},\\
&\langle
\downarrow,\theta,\phi
|
\sigma_{-}
|
\downarrow,\theta,\phi
\rangle=-\sin\theta\rme^{-i\phi},\\
&\langle
\downarrow,\theta,\phi
|
\sigma_{-}
|
\uparrow,\theta,\phi
\rangle=2
\left[
\cos\frac{\theta}{2}
\right]^2
\rme^{-i\phi},\\
&\langle
\uparrow,\theta,\phi
|
\sigma_{-}
|
\downarrow,\theta,\phi
\rangle=-2
\left[
\sin\frac{\theta}{2}
\right]^2
\rme^{-i\phi}.
\end{aligned}
\ee

The product $\vn{L}^{\alpha}\cdot\vn{\sigma}$ in Eq.~\eqref{eq_soi_ham} may
be rewritten 
as
$\vn{L}^{\alpha}\cdot\vn{\sigma}=L^{\alpha}_z\sigma_z+[L^{\alpha}_+\sigma_-+L^{\alpha}_-\sigma_+]/2$.
Therefore, we need to multiply the matrix elements in Eq.~\eqref{eq_app_ksi_R}
with the angular factors in Eq.~\eqref{eq_sigma_z}, 
Eq.~\eqref{eq_sigma_plus}, and Eq.~\eqref{eq_sigma_minus}
in order to obtain the matrix elements of the SOI Hamiltonian Eq.~\eqref{eq_soi_ham}
in the basis set of the MLWFs without SOI.
The resulting matrix elements are given in Eq.~\eqref{eq_v_upup},
Eq.~\eqref{eq_v_downdown}, 
Eq.~\eqref{eq_v_updown}, and Eq.~\eqref{eq_v_downup} in the main text.

\bibliography{lasintorafm}

\end{document}